\documentclass[conference]{IEEEtran}
\IEEEoverridecommandlockouts
\usepackage{cite}
\usepackage{amsmath,amssymb,amsfonts}
\usepackage{algorithmic, booktabs}
\usepackage{graphicx}
\usepackage{textcomp}
\usepackage{xcolor}
\def\BibTeX{{\rm B\kern-.05em{\sc i\kern-.025em b}\kern-.08em
    T\kern-.1667em\lower.7ex\hbox{E}\kern-.125emX}}
\begin{document}

\title{Evaluating Atypical Gaze Patterns through Vision Models: The Case of Cortical Visual Impairment}

\author{
  \IEEEauthorblockN{
    Kleanthis Avramidis\IEEEauthorrefmark{1},\quad
    Melinda Y. Chang\IEEEauthorrefmark{2},\quad
    Rahul Sharma\IEEEauthorrefmark{3},\quad
    Mark S. Borchert\IEEEauthorrefmark{2},\quad
    Shrikanth Narayanan\IEEEauthorrefmark{1}
  }
  \IEEEauthorblockA{
    \IEEEauthorrefmark{1}Signal Analysis and Interpretation Lab,
    University of Southern California, Los Angeles, CA 90089
  }
  \IEEEauthorblockA{
    \IEEEauthorrefmark{2}Roski Eye Institute,
    University of Southern California, Los Angeles, CA 90089
  }
  \IEEEauthorblockA{
    \IEEEauthorrefmark{3}Amazon\thanks{This work was done when Rahul Sharma was at USC.}
  }
}\vspace{-0.2cm}

\maketitle
%
%
%
%
%

\begin{abstract}
    A wide range of neurological and cognitive disorders exhibit distinct behavioral markers aside from their clinical manifestations. Cortical Visual Impairment (CVI) is a prime example of such conditions, resulting from damage to visual pathways in the brain, and adversely impacting low- and high-level visual function. The characteristics impacted by CVI are primarily described qualitatively, challenging the establishment of an objective, evidence-based measure of CVI severity. To study those characteristics, we propose to create visual saliency maps by adequately prompting deep vision models with attributes of clinical  interest. After extracting saliency maps for a curated set of stimuli, we evaluate fixation traces on those from children with CVI through eye tracking technology. Our experiments reveal significant gaze markers that verify clinical knowledge and yield nuanced discriminability when compared to those of age-matched control subjects. Using deep learning to unveil atypical visual saliency is an important step toward establishing an eye-tracking signature for severe neurodevelopmental disorders, like CVI.
\end{abstract}

\section{Introduction}
\vspace{-0.1cm}
Human behavior presents a complex interplay influenced by social surroundings, the environment, and an individual's cognitive or biomedical conditions. There is hence an increasing interest in studying behavioral markers that could indicate neurological diseases (e.g., Alzheimer's disease~\cite{ehrlich2021association}), mental illness (e.g., depression~\cite{arevian2020clinical}), developmental disorders (e.g., Autism Spectrum Disorder~\cite{guha2016computational}) or affective patterns (e.g., mood~\cite{appelhans2006heart}, stress~\cite{hjortskov2004effect}). Aside from the profound implications to the quality and length of human life, such conditions incur a heavy financial burden for testing and treatment. The investigation of behavioral markers could thus assist specialists in both interpreting those conditions and diagnosing them with reduced costs. Behavioral Signal Processing (BSP)~\cite{bone2017signal} and Artificial Intelligence (AI) technologies seek to quantify subtle nuances of behavioral signals that may serve as indicators of underlying cognitive or biomedical conditions. Various modalities have been researched to that end, with the most prominent being speech and paralinguistic patterns~\cite{narayanan2013behavioral}, facial expressions~\cite{guha2016computational}, movement and physiology, as quantified through specialized biosensors~\cite{hjortskov2004effect, frank2023wearable}.

In this paper, we apply this premise and use eye tracking technology to study Cortical Visual Impairment (CVI), one of the leading causes of pediatric visual impairment worldwide~\cite{chang2020advances}. Yet, CVI lacks standardized, evidence-based methods that interpret its impacted visual characteristics. Eye tracking is a favorable behavioral modality, as it has been previously utilized to quantify atypical gaze patterns to predefined stimuli~\cite{vidal2012wearable}. By scrutinizing gaze patterns, experts can gain a deeper understanding of how clinical populations perceive their environment~\cite{skaramagkas2021review}. Notably, people in the ASD spectrum report reduced attention to social stimuli such as the human face, voice and hand gestures~\cite{sasson2011brief, wang2015atypical}, whereas people with diagnosed depression show attentional biases to emotional faces~\cite{duque2015double}. In our case, eye tracking could quantify the visual function of CVI patients on images of higher-order attributes such as social interaction and texture~\cite{manley2022assessing, bauer2023deficits}. However, assessing human visual saliency requires a model for efficient extraction of saliency for the stimuli used for diagnosis. This is typically done through expert annotations, however these are time-consuming and prone to individual biases. Computational models~\cite{kummerer2016deepgaze,midas} have thus emerged to ease this process, yet they rarely capture high-level semantics.

In the following, we describe a machine learning-driven approach to create custom maps of visual saliency by prompting semantic and qualitative characteristics of CVI to multimodal vision models. We first outline a set of saliency markers related to CVI and an experiment setup to study those through measures of eye tracking. We engineer suitable prompts for attributes of interest and show that the derived saliency maps offer nuanced insights that are clinically relevant.

\section{Cortical Visual Impairment}

Cortical, or cerebral, visual impairment (CVI) is a neurological condition that occurs due to damage or injury to the visual pathways in the brain, resulting in a range of visual deficits~\cite{jan1987behavioural}. Common causes of CVI include prematurity with periventricular leukomalacia, hypoxic-ischemic encephalopathy, trauma, hydrocephalus, metabolic or genetic disorders~\cite{chang2020advances}. CVI is the leading cause of pediatric visual impairment in developed countries, yet there is no standardized method of quantifying the impacted visual characteristics~\cite{chang2021methods}, and no evidence-based treatment option is available.

Despite the prevalence of CVI, the precise definition remains a topic of debate in the literature. Traditionally, the diagnosis of CVI required decreased visual acuity and/or visual field defects~\cite{whiting1985permanent}. Most children diagnosed using this definition had profound visual impairment or blindness at diagnosis. More recently, the term CVI has been used more broadly to include children with deficits of higher-order visual function (such as recognition of abstract objects or faces), with or without intact visual acuity and visual fields. This has led to the consensus definition proposed by Sakki et al.~\cite{sakki2017there} of CVI as ``a verifiable visual dysfunction which cannot be attributed to disorders of the anterior visual pathways or any potentially co-occurring ocular impairment."

Still, several unifying characteristics among children with CVI have been reported in the literature for decades, mainly based on qualitative descriptions. These include reduced contrast sensitivity~\cite{good2012spatial} and spared color discrimination~\cite{cohen2005visual}, selective sparing or limitation of motion processing~\cite{pamir2021neural}, difficulties with visual crowding or complexity~\cite{manley2022assessing}, specific deficits in recognizing abstract objects or faces (e.g. prosopagnosia)~\cite{bauer2023deficits}, difficulties with visually-guided orientation in space (topographic agnosia)~\cite{dutton1996cortical}, and variability in visual function based on individual and environmental factors~\cite{jan1987behavioural}. Since most children with CVI have developmental delays that impair communication, identifying these deficits relies on assessment of visual behavior. Eye tracking is an attractive option to that end because, in addition to being objective and quantitative, the tracking protocols may be designed to assess a multitude of lower- and higher-order visual characteristics.\vspace{-0.1cm}

\begin{figure}
    \centering
    \includegraphics[scale=0.255]{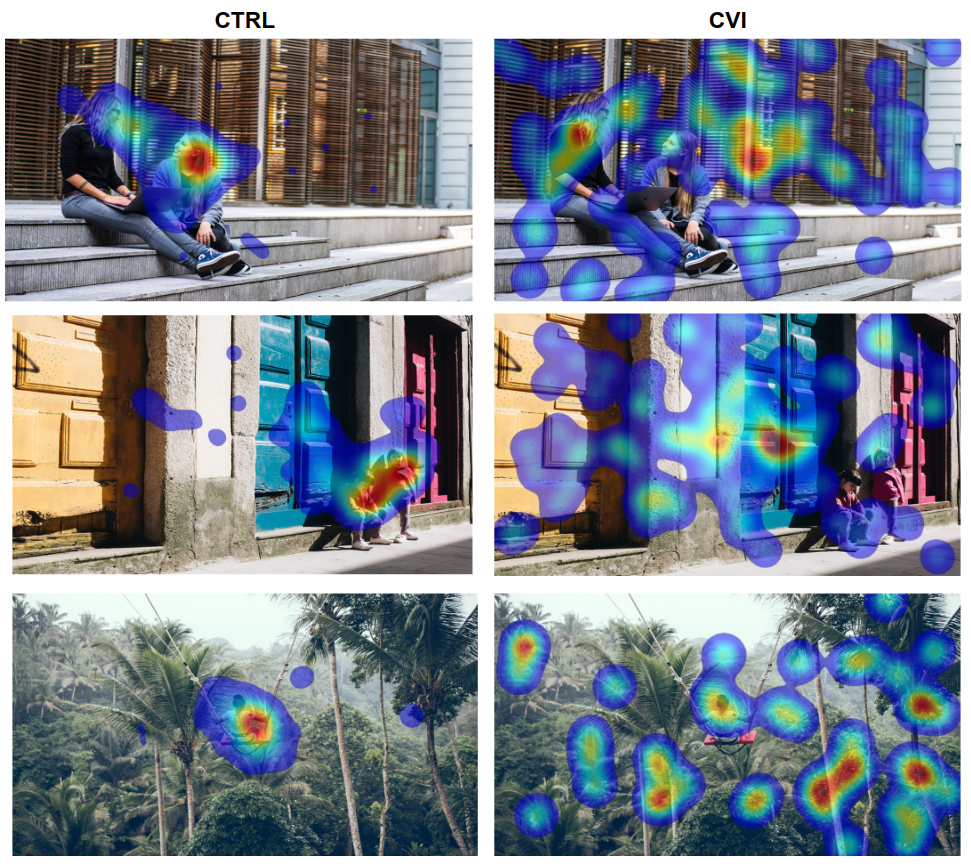}
    \vspace{-0.2cm}
    \caption{Fixation density maps from aggregate controls (CTRL) and children with CVI on sample experimental stimuli. The heatmap shows fixation density.}
    \vspace{-0.3cm}
    \label{fig:fixation-density}
\end{figure}

\section{Model-based Visual Saliency}
Models of visual saliency traditionally include a set of handcrafted features extracted from the visual scene (images) through complex image processing techniques. Such features include color patterns or opponencies (e.g., red against green), edges or orientation patterns that have been shown to be relevant for human vision~\cite{itti1998model}. With the emergence of deep learning, there are more high-level attributes to account for, including shapes, depth and other human-inspired patterns. Neural network models like DeepGaze~\cite{kummerer2016deepgaze} have been trained on human fixations to predict an average image saliency map.

However, these bottom-up approaches are based solely on stimulus features. On the other hand, top-down human saliency takes into account human conditions, past experiences and intent to estimate fixation patterns~\cite{tanner2019top}. Hence, CVI patients would be expected to diverge from baseline fixation estimations due to the associated visual impairment. Moreover, conditions like CVI require the extraction of saliency features that are challenging to extract with those methods, e.g., maps based on specific object categories, or based on non-material attributes like social interaction, motion, and complexity.

Multimodal learning, particularly vision-language models, have been proposed to ground the visual modality to text descriptions. CLIP~\cite{radford2021learning} is among the first efficient multimodal models which include two unimodal encoders for visual and textual inputs respectively. CLIP is trained by minimizing the latent distance of paired images and text descriptions, whereas maximizing unpaired sample distances using the InfoNCE~\cite{oord2018representation} contrastive objective. Here we adopt SegCLIP~\cite{luo2023segclip}, a recent variant of CLIP which is specifically trained for text-driven image segmentation. This task is particularly suited for our study since the model output is a saliency map that corresponds to the segmentation of the prompted attributes.

\textbf{Prompt Engineering}\, SegCLIP incorporates a dual-encoder architecture containing a text and image encoder. The authors suggest three losses, including a reconstruction loss that trains the generated segmentation output and a contrastive loss that aligns text and image inputs. We use the pre-trained model in its inference mode and provide image inputs (our stimuli) along with free-form text descriptions of the attributes of interest (prompts). The model then aligns the two inputs and outputs a segmentation map that highlights the corresponding attribute in the image (hence, a saliency map). While SegCLIP is trained primarily on object-centric prompts, authors show in \cite{luo2023segclip} that the model can generalize to action attributes like flying, driving, etc. Here we investigate the extent to which it can generalize to qualitative attributes, such as complexity or social concepts, and we engineer prompts to better reflect the nuances of the CVI characteristics.

\begin{figure*}
    \centering
    \includegraphics[scale=0.27]{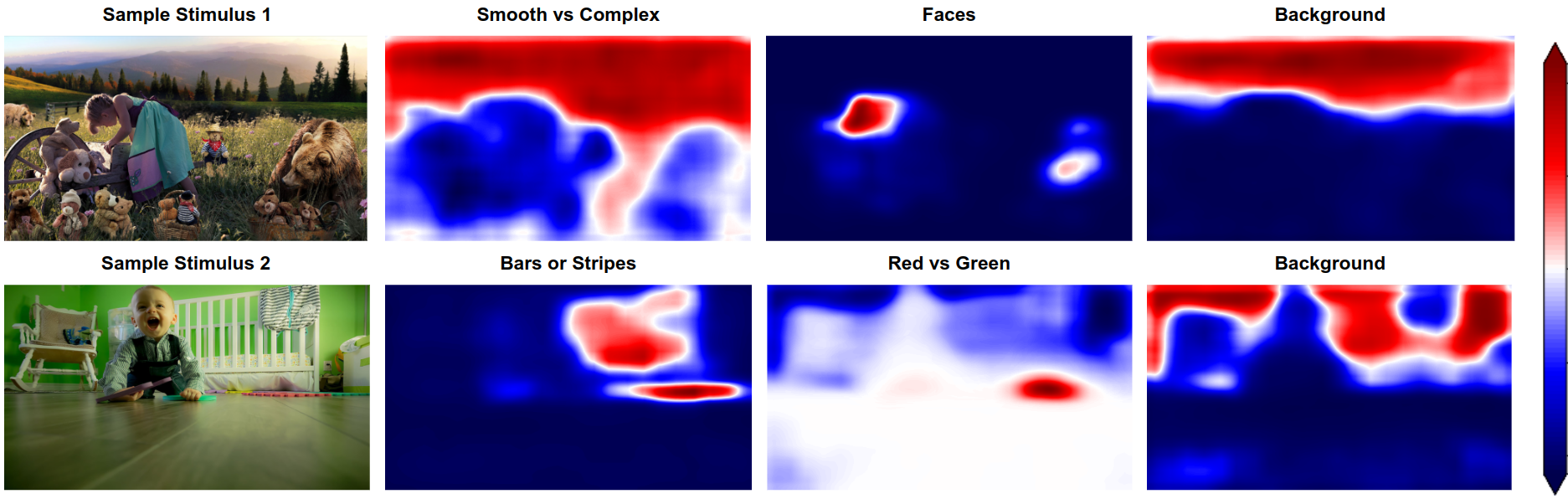}
    \vspace{-0.1cm}
    \caption{Visual saliency maps with corresponding text \textit{prompts}, as extracted from the original stimuli (on the left) using SegCLIP. Red is higher.}
    \vspace{-0.4cm}
    \label{fig:segclip-maps}
\end{figure*}

\section{Experimental Setup}
\subsection{Study Design and Participants}
We prospectively recruited 42 children between 12 months to 12 years of age with a diagnosis of CVI. The diagnosis was made during a routine eye exam by a pediatric neuro-ophthalmologist, based on observed reduced or worse than expected visual acuity in children, given the severity of any ocular co-morbidity. Additionally, recruited children were required to have a known neurologic risk factor for CVI (e.g., prematurity with periventricular leukomalacia). We included children who had visual acuity sufficient to track images on a computer monitor and excluded those with photosensitive epilepsy and any ocular cause for decreased vision, except for mild atrophy. We also excluded children with abnormalities such as oculomotor apraxia that would prevent us from making inferences based on eye movements.

We also recruited 29 age-matched control subjects using a web-based recruitment service. Controls had no history of any neurologic or ophthalmologic condition, and underwent an eye exam to confirm normal visual acuity and ocular motility. 
The study was approved by the Children’s Hospital Los Angeles/University of Southern California Institutional Review Board (IRB) and adhered to the tenets of the Declaration of Helsinki and the US Health Insurance Portability and Accountability Act of 1996. Informed consent was obtained from the parent or legal guardian of all participants.

Eye tracking was performed using the SR Research EyeLink 1000 (Ottawa, Canada)\footnote{https://www.sr-research.com/eyelink-1000-plus/} desktop remote eye tracker, recording at 500 Hz. Participants were seated 60 cm from a computer monitor, wearing their habitual spectacles. The recording session began with 3-point calibration and proceeded for 10 minutes while the participants watched a series of still images and videos interspersed with stimuli for psychophysical tests~\cite{chang2023comparison,chang2021validity}. In this paper we focus on still images, including pictures of landscapes, people, animals, toys, and food, that were selected to span the range of characteristics believed to be affected in CVI (e.g., color, contrast, complexity, orientation, and human interaction).  Both realistic and cartoon images were included. Each image was shown for a total of 2 seconds.\vspace{-0.1cm}

\subsection{Eye Tracking Signal Analysis}
The eye-tracking software outputs gaze coordinates, pupil size, and sample-level annotations for detected fixations (i.e., periods during which gaze remains at a particular location), saccades (i.e., rapid eye movements between fixations), and eye blinks. For an eye movement to be considered fixation, it has to remain stable (at most 0.1 deg) for at least 100ms. For an eye movement to be considered saccade, it should have a velocity of at least 30 deg/s at an amplitude change of at least 0.1 deg. Here we focus on the gaze trajectory projected on stimulus image coordinates, which we partition into fixation and saccade segments. We discard fixations shorter than 50 ms~\cite{albert2022measuring} and further remove any fixations positioned partially (at least 20\%) out of the stimulus image range. Lastly, we averaged the respective traces of the two eyes when both were available. To facilitate our statistical analysis of CVI markers, we consider the resulting trajectories per image trial and compute the average saliency of the fixation traces.
\subsection{Statistical Analysis of Saliency}
We evaluate the aforementioned visual characteristics of CVI by comparing the extracted saliency features across groups. The saliency maps are extracted from SegCLIP using free language prompts. For instance, depth was approached using prompts like ``\textit{depth}", ``\textit{background}" versus ``\textit{foreground}", or ``\textit{far away}" versus ``\textit{front}". Similarly, complexity was induced with prompts such as ``\textit{complex scene}", or ``\textit{complexity}" versus ``\textit{smoothness}". The generated grayscale maps were then smoothed with a gaussian filter and normalized to [0, 255]. We carried out a series of non-parametric Mann-Whitney tests to unveil significant differences between the two groups, accompanied by Cohen's $d$ effect size and permutation testing ($p_\text{perm}$). We consider our results significant if $p\equiv p_{MW}<0.01$.

\textbf{Differential Saliency} In the previous examples, the use of \textit{versus} indicates differential saliency between opposite attributes of interest. This is done by extracting the saliency maps using two separate prompts, and then getting their subtraction (note that order is important) as the final map. We manually observed that this technique enhances the precision to attributes of interest and regularizes the effect of CVI patients producing significantly fewer fixations per trial (CTRL: $3.8$, CVI: $2.2, p<10^{-11}, d=2.68, p_\text{perm}=0.0002$), with those being scattered from the image center (center bias -- CTRL: $217$, CVI: $201, p<10^{-12}, d=3.17, p_{\text{perm}}=0.0002$).

\begin{figure*}
    \centering
    \includegraphics[scale=0.3]{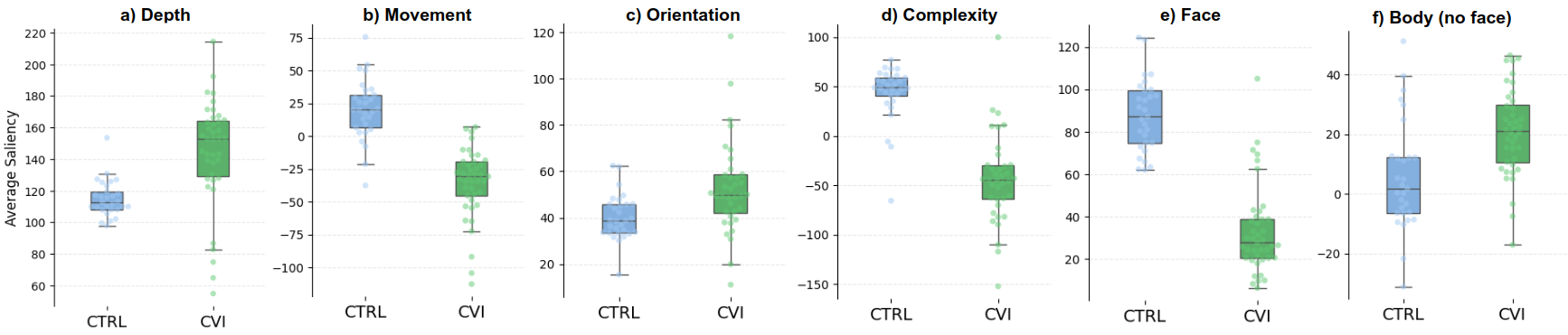}
    \vspace{-0.6cm}
    \caption{Average fixation saliency aggregated per subject, for attributes of interest. CVI subjects a) are sensitive to depth information, as they fixate less on foreground elements, b) show preference toward still objects, encoded with lower movement saliency, c) focus more on directional patterns, d) fixate on smooth rather than complex/crowded parts, e) avoid directly looking at faces, and f) prefer non-social human characteristics compared to faces.}
    \vspace{-0.3cm}
    \label{fig:avg-saliency}
\end{figure*}

\begin{figure}
\centering
    \includegraphics[scale=0.185]{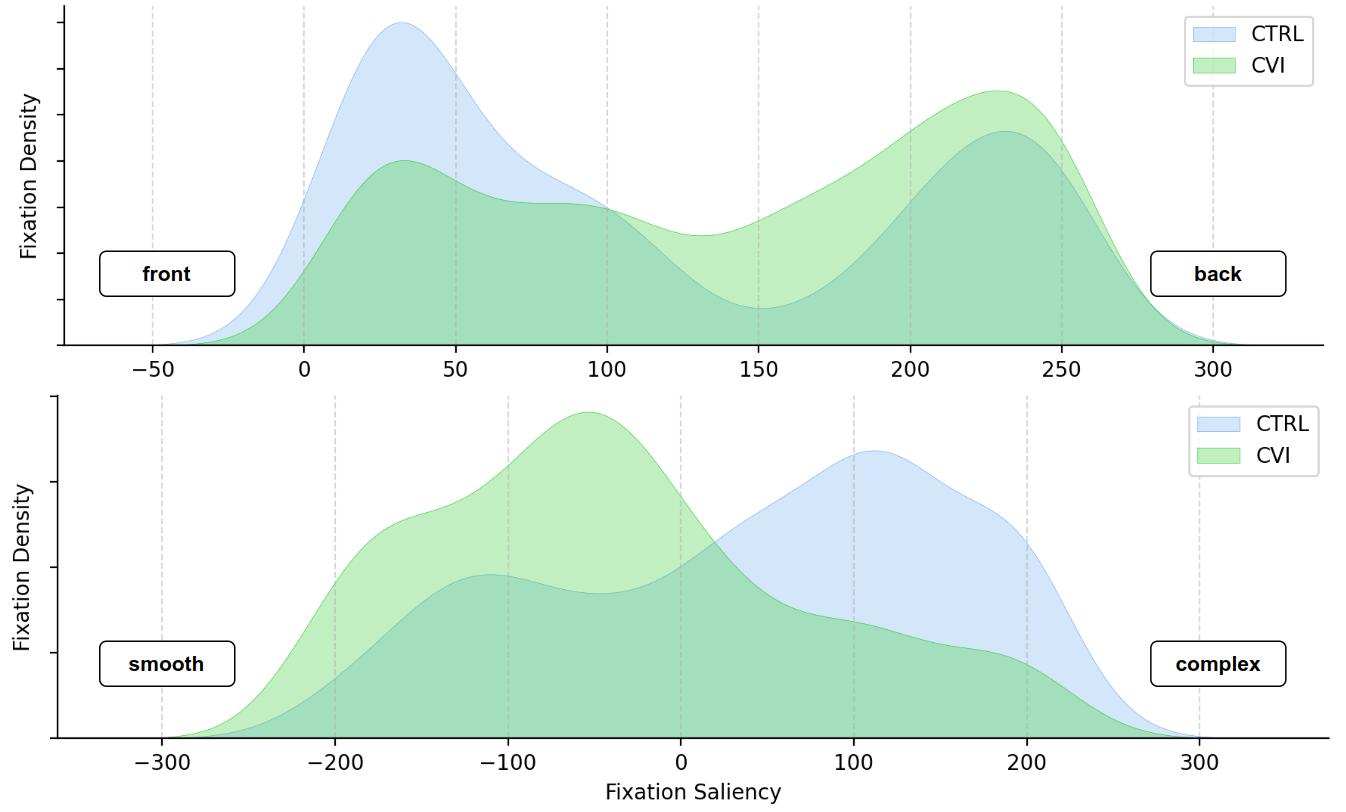}\\
    \vspace{-0.1cm}
    \caption{Fixation density for \textit{depth} (up) and \textit{complexity} (down) saliency, reflecting the atypical gaze patterns of CVI patients in our experiment.}
    \label{fig:stat-plots}
    \vspace{-0.4cm}
\end{figure}

\section{Results \& Discussion}

We first verify that CVI patients demonstrate atypical visual saliency compared to the matched controls. We use DeepGaze~II~\cite{kummerer2016deepgaze} to create saliency maps that estimate typical human gaze patterns for our stimuli. Indeed, CVI patients report an average fixation saliency of 53, compared to 158 for controls ($p < 10^{-17}, d = 2.12, p_{\text{perm}} = 0.0002$), which verifies the assumption of atypical gaze patterns in CVI. In the following we analyze each of the prominent characteristics of CVI in terms of average fixation saliency measured across groups, and summarize the results in Figure~\ref{fig:avg-saliency}.

\vspace{0.1cm}
\textbf{Depth and Background} For this experiment we extracted depth saliency maps by prompting for \textit{depth}, with foreground information denoted with higher saliency. As shown in Figure~\ref{fig:avg-saliency}a, patients with CVI focus significantly more to distant information in the image. In specific, the computed average is 107 for the control group, compared to 140 for CVI ($p < 10^{-5}, d = 1.18, p_{\text{perm}} = 0.0002$). This result is also verified using depth estimation maps, extracted using MiDaS~\cite{midas}. By plotting the corresponding fixation density (Figure~\ref{fig:stat-plots}) we observe that the depth bias comes primarily from CVI tendency to focus less on foreground than more to background saliency. Despite this tendency, CVI subjects need more time to produce a valid background fixation ($p=0.0005, d = 0.70, p_{\text{perm}} = 0.0002$), while their overall latency to salient fixations does not differ ($p=0.372$).

\vspace{0.1cm}
\textbf{Orientation and Motion} Consistent with the hypothesis, CVI subjects focus on \textit{still} rather than \textit{moving} objects in an image ($p < 10^{-9}, d = 2.28, p_{\text{perm}} = 0.0002$), as shown from the differential saliency reported in Figure~\ref{fig:avg-saliency}b. Regarding orientation information, SegCLIP outputs the most representative maps when prompted with \textit{bars or stripes}, as compared to more abstract prompts like \textit{orientation}, or \textit{direction} and \textit{directional patterns}. This is expected as the model performs better on object-centric prompts. As seen in Figure~\ref{fig:avg-saliency}c, CVI patients seem to fixate more to such patterns than the control group ($p=0.006, d = 0.81, p_{\text{perm}} = 0.004$).

\vspace{0.1cm}
\textbf{Contrast and Complexity} For this experiment we tested the assumption that CVI produces functional challenges regarding scenes of high complexity, with a generally reduced contrast sensitivity~\cite{good2012spatial}. As shown in Figure~\ref{fig:avg-saliency}d, CVI subjects fixate significantly less on \textit{complex} (positive saliency) rather than \textit{smooth} (negative saliency) parts of the corresponding stimuli ($p < 10^{-8}, d = 2.12, p_{\text{perm}} = 0.0002$). This symmetric difference is evident also from the fixation density produced by the 2 groups (Figure~\ref{fig:stat-plots}). Similarly, testing on a subset of stimuli characterized by \textit{texture} reveals that CVI subjects tend to avoid those regions, reporting an average saliency of 98 compared to 110 for the controls ($p=0.003, d = 0.66, p_{\text{perm}} = 0.002$).

\vspace{0.1cm}
\textbf{Faces and Social Stimuli} CVI is assumed to distract gaze from social stimuli. For this experiment we isolate multiple groups of stimuli images corresponding to different concepts of human interaction: \textit{faces}, \textit{hands}, \textit{eyes}, etc. In this case, SegCLIP was prompted with single-object descriptions to yield representative saliency maps. Prompting with abstract concepts like \textit{human interaction}, \textit{communication} or \textit{emotion} did not produce consistent maps for our stimuli.  The results of these experiments are summarized in Figure~\ref{fig:avg-saliency}e-f. CVI subjects are shown to fixate less than average on human \textit{faces} ($p < 10^{-10}, d = 2.86, p_{\text{perm}} = 0.0002$), while also the latency of these fixations is longer ($p < 10^{-4}, d = 0.87, p_{\text{perm}} = 0.0002$). No difference emerges when it comes to fixations on \textit{eyes} or \textit{mouth} ($p>0.75$). Interestingly though, as shown in Figure~\ref{fig:avg-saliency}f, children with CVI tend to focus more on the non-social human characteristics ($p=0.001, d = 0.59, p_{\text{perm}} = 0.0016$). An additional within-group difference is that controls produce longer fixations at higher saliency (pearson $\rho=0.21$, $p=0.001$), whereas CVIs report uniform durations ($p=0.675$).

\section{Conclusion}

A novel, machine learning-based approach to create maps of visual saliency was proposed, focusing on the assessment of qualitative markers of eye-tracking activity in children with CVI. To that end, we leveraged the robustness of large vision models in segmenting image patches conditioned on textual descriptions of CVI characteristics. The availability of those saliency maps assisted us in verifying and reinforcing our understanding of a challenging neurological impairment with no standardized diagnostic methods. In the future, the proposed approach could be used in designing individualized interventions and assessing their efficacy in clinical trials.

\bibliographystyle{unsrt}
\bibliography{refs.bib}
\end{document}